\documentclass[11pt]{article}
\usepackage{UF_FRED_paper_style}

\usepackage{lipsum}  

\title{Principio de Contrastibilidad, Geometría y Sistemas de Conocimiento
}

\author{Carlos Desa\\
    \href{mailto:firstauthor@ufl.edu}{\texttt{170735@unsaac.edu.pe}} 
    }
    
\date{}

\begin{document}
{\setstretch{.8}
\maketitle
\begin{abstract}
There is a contemporary trend toward geometrizing all mathematical theories, as proposed by the Langlands program, and, by extension, physical theories as well. Within this paradigm, it becomes possible to represent physical objects as formal geometric entities. This opens the door—within the framework of logical empiricism-pragmatism and constructive realism—to the development of a new philosophical approach grounded in a logical-mathematical perspective. The aim is to integrate both the physics and metaphysics of the objects of the world into a unified system of knowledge, based on the geometric and physical interpretations of their mathematical representations. This article is presented as a foundational step in the development of that approach.\\

%
\noindent
\end{abstract}
}

%
\section{Introducción}
La naturaleza de las matemáticas, y por ende de las teorías científicas, ha intrigado a científicos y filósofos desde los orígenes de ambas disciplinas. La efectividad descriptiva de la física constituye un punto clave en esta búsqueda. En particular, la corriente contemporánea de geometrización de la física~\cite{schuller2017lectures,doring2015new,carcassi2022geometric}, dado que el programa de langlands abre la  posibilidad de geometrizar cualquier teoría matemática~\cite{gaitsgory2024proof}, y por tanto cualquier teoría física, se busca reformular las teorías de la física a partir de principios físicos simples~\cite{ball2017quantum}. Esta corriente junto con el principio de contrastabilidad —según el cual toda afirmación sobre los objetos del mundo debe poder ser contrastada de algún modo, ya sea empíricamente, mediante la experiencia, o pragmáticamente, en función de su efectividad explicativa dentro de un marco más general de comprensión de los objetos del mundo (Se entiende por “objetos del mundo” no solo los objetos en sí, sino también sus propiedades y sus relaciones con otros objetos)-, ofrece un nuevo paradigma para abordar esta cuestión. Este enfoque abre el camino hacia una nueva metafísica, orientada a resolver el problema de la naturaleza de las teorías científicas. El objetivo es alcanzar una comprensión general de las representaciones geométricas de los objetos del mundo y de sus implicaciones filosóficas, dentro de un marco lógico-matemático más amplio. El presente articulo se presenta como un punto de partida, bajo las concepciones mencionadas, encaminado en la búsqueda de esta respuesta.\\
Para este fin, es importante reconocer la capacidad innata del ser humano para identificar patrones, organizarlos y desarrollado un lenguaje simbólico, puede representarlos. Desde el realismo-constructivo~\cite{tcytcarev2019constructive} y el empirismo-pragmatismo  lógico~\cite{neuber2015realistic}, este lenguaje simbólico —en especial el formal, lógico-matemático— adquiere sentido dentro de las teorías científicas. También debe considerarse que, bajo ciertas condiciones, es razonable confiar en nuestras percepciones del mundo y en el sentido común.\\
En el primer capítulo se ofrece una explicación del origen de la lógica dentro del marco conceptual previamente expuesto. El segundo capítulo está dedicado al concepto de lenguaje ideal, entendido como un lenguaje sin ambigüedades. Finalmente, en el tercer capítulo se abordan, a partir de estas consideraciones previas, la geometrización de la física y algunas pistas relevantes que podrían orientar el desarrollo de una nueva filosofía en este contexto.

\section{Lógica}
Los objetos del mundo y sus cambios existen objetivamente. Parte de este mundo somos los humanos, quienes perciben parte de estos objetos a través de sensaciones (mundo sensible). Dada nuestra capacidad innata de reconocimiento de patrones, con la inducción estadística, identificamos en los objetos del mundo dichos patrones. Puesto que los objetos del mundo son de una manera y no de otra, y por tanto consistentes, en el marco del empirismo lógico, haciendo uso del lenguaje formal, es posible representar estos objetos del mundo, aproximadamente, con objetos formales. \\
Por otro lado, existe la necesidad de suposiciones lógicas sobre el mundo no sensible, sus aspectos esenciales y para el funcionamiento de los sistemas de conocimiento (no-triviales). Por tanto, para un entendimiento más completo y por tanto más acorde a los objetos del mundo. Es necesaria una lógica pragmática, que por su efectividad explicativa toma una forma, siendo su efectividad relativa a los objetos del mundo. \\
Este conjunto de reglas, pragmáticas y empíricas, constituye la base de los sistemas de conocimiento, su lógica inherente. Estas reglas, que buscan eliminar en lo posible las inconsistencias, la trivialización del sistema y las ambigüedades, es a lo que se denominara como lógica. \\
En el lenguaje, dentro de sus arbitrariedades y sus "juegos", se busca que, en cuanto al sentido de los símbolos —y por tanto a lo que hacen referencia en el mundo—, exista una relación directa: el significado es la referencia. Buscando una correspondencia uno a uno entre los símbolos (objetos formales) y los objetos del mundo. \\
Es necesario considerar también que, puesto que el lenguaje, y por tanto nuestro mundo, son limitados, hay una parte del mundo —el noúmeno— que no tiene una correspondencia empírica en el lenguaje. En este sentido, los objetos del mundo tendrían una correspondencia inyectiva con el lenguaje. Entonces, en la necesidad e cubrir estos "huecos" de comprensión, se da lugar a una lógica  netamente pragmática. 
%
\section{Lenguaje No Ambiguo}
En el sentido más estricto del lenguaje, dada su relación inyectiva con el mundo, se busca el conjunto de símbolos que permita describir los objetos del mundo, de manera que se eliminen, en lo posible, las ambigüedades naturales del lenguaje, y por tanto se obtenga un significado claro y conciso que remita al objeto, perfeccionar el sentido, y por tanto la referencia.\\
En este lenguaje ideal, el sentido de los símbolos lógicos sería estrictamente sintáctico —lógica formal—, siendo las bases de estos sistemas enunciados sintéticos. Sin embargo, es necesaria la semántica para la definición y también para la interpretación, y por consiguiente para el entendimiento de estos sistemas.\\
Entendido aquello, es pues el campo de la lógica formal o matematizada y la matemática, en conjunto lógica-matemática, el lenguaje que cumple con tal fin.

\section{Principios y Teorías}
La lógica-matemática constituyen el estudio de objetos formales definidos lógicamente. Así, la matemática se ocupa del análisis de estos objetos y de sus relaciones con otros. En un sentido lógico, se establecen vínculos entre los objetos del mundo -dado que son comparables y por tanto medibles, cuantificables- y estos objetos formales, intentando, de manera aproximada, captar las cualidades de los primeros y, por tanto, cómo se relacionan con el resto del mundo, siendo esta una representación ideal del mundo. De este modo, los enunciados matemáticos adquieren un sentido físico, dada su referencia real. \\
En este sentido, podríamos referirnos a tres aspectos esenciales de las teorías científicas, y, en su totalidad, de un sistema de conocimiento. En él deben contenerse tres tipos de enunciados: observacionales, teóricos y metafísicos (suposiciones metafísicas). \\
Los enunciados observacionales son puramente empíricos, y dentro de un marco teórico, son explicados como efectos directos de los enunciados teóricos. Estos últimos son, en parte, empíricos —al estar directamente relacionados con los observacionales— y en parte pragmáticos, ya que se fundamentan de a acuerdo a su eficiencia explicativa. \\
En el marco de todo un sistema de conocimiento, incluso en el  marco teórico, se hallan los enunciados metafísicos, los cuales, por su naturaleza, en un sentido puramente pragmático, trascienden los métodos de la ciencia, siendo pues necesarios como parte de los fundamentos de los sistemas de conocimiento, fuertemente relacionados con las resultados de la ciencia, enriquecidos y diseccionados por estos, en este sentido, pertenecientes a una filosofía científica~\cite{romero2019filosofia}.\\
Los objetos del mundo presentan simetrías, las cuales pueden ser representadas topo-geométricamente. Y dado el teorema de Noether, estas implican invariantes: principios de conservación que, en un sentido más amplio, constituyen principios físicos. \\
Dichos invariantes —y, por lo tanto, las simetrías— pueden ser de naturaleza ontológica o no; sin embargo, en ambos casos deben entenderse su manifestación epistemológica. Esto se debe a que las simetrías, y por ende las invariancias, se manifiestan tanto en los objetos del mundo tal como son en sí mismos o como en la manera en que solo son percibidos (solamente subjetivas). En cualquier caso, estas simetrías —y las invariancias asociadas— se expresan siempre a través de nuestra percepción de los objetos del mundo. \\
Se entiende por simetrías no ontológicas, también denominadas epistemológicas netamente, aquellas que surgen al estudiar los objetos del mundo tal como se manifiestan en las estructuras mentales, sistemas de referencia inerciales. Estas simetrías, origen de cambios en los objetos del mundo no corresponden a cambios  en el objeto en sí, sino a cambios en como es percibido, en el sujeto, y por tanto, subjetivo arbitrariamente.\\
Sin embargo, no importa cómo sea esta estructura mental o sistema de referencia particular, sino cómo es el objeto en si mismo. De este modo, se manifiestan simetrías objetivas, independientemente del observador -principio de relatividad~\cite{einstein1905elektrodynamik}-.
Sea entonces que un marco teórico establece de manera formal -representación- la relación de los objetos en si mismos y en como son percibimos.
Podemos afirmar entonces que una teoría científica, la cual se fundamenta en principios físicos que admiten una interpretación geométrica, donde se identifica a los objetos del mundo con entes formales, específicamente los de la topología geométrica- es decir la teoría tiene una estructura geométrica~\cite{schuller2017lectures}-, acompañada por un conjunto de suposiciones metafísicas, constituye un sistema de conocimiento completo sobre los objetos del mundo. Dicho sistema es de carácter analítico-deductivo.\\
Sin embargo, a medida que los objetos del mundo se alejan de su representación idealizada dentro del modelo, estos esquemas teóricos tienden a perder su eficacia predictiva.\\
También debe entenderse la práctica científica, y sus aspectos filosóficos, como un proceso constante de perfeccionamiento del lenguaje, de su sentido y por tanto de su referencia, en función de su correspondencia con los objetos del mundo. Estas doctrinas filosóficas (específicamente la metafísica) adquieren un carácter netamente pragmático: su valor radica en su capacidad de ser contrastadas y en su efectividad dentro de las prácticas científicas.
\section{Conclusiones}
La tendencia contemporánea de reformular las teorías físicas mediante postulados más simples, representando los objetos del mundo como entidades formales en el marco de la topología geométrica, junto con el programa de Langlands, plantea un nuevo camino en la filosofía para una comprensión renovada de los objetos del mundo (su física y metafísica). En este contexto, dentro del marco de la lógica-matemática y del constructivismo-realista, se abre la posibilidad de construir una teoría general de los objetos del mundo a partir de sus interpretaciones geométricas.
%

\medskip

\bibliography{references.bib} 

\newpage




\end{document}